\documentclass[a4paper,11pt]{article}
\pdfoutput=1 

\usepackage{jheppub} 

\usepackage[T1]{fontenc} 

\usepackage{slashed}
\usepackage{extarrows}

\title{\boldmath Elastic proton-proton scattering at LHC energies in holographic QCD}

\author[a,b,1]{Wei Xie,\note{Corresponding author.}}
\author[b,c]{Akira Watanabe}
\author[c]{and Mei Huang}

\affiliation[a]{College of Science, China Three Gorges University,\\Yichang 443002, People's Republic of China}
\affiliation[b]{Institute of High Energy Physics and Theoretical Physics Center for Science Facilities, Chinese Academy of Sciences,\\Beijing 100049, People's Republic of China}
\affiliation[c]{School of Nuclear Science and Technology, University of Chinese Academy of Sciences,\\Beijing 100049, People's Republic of China}

\emailAdd{xiewei@ctgu.edu.cn}
\emailAdd{akira@ihep.ac.cn}
\emailAdd{huangmei@ucas.ac.cn}

\abstract{
The cross sections of the high energy proton-proton scattering are studied in a holographic QCD model, focusing on the Regge regime.
In our model setup, the involved nonperturbative partonic dynamics is described by the Pomeron exchange, which is realized by applying the Reggeized spin-2 particle propagator together with the proton gravitational form factor obtained from the bottom-up AdS/QCD model. Our model includes three adjustable parameters which are to be fitted by experimental data. We show that both the resulting differential and total cross sections are consistent with data, including the ones recently measured at $\sqrt{s}=13$~TeV by the TOTEM collaboration at the LHC. Our results imply that the present framework works well in the considered TeV scale, and further applications to other high energy scattering processes, in which the involved strong interaction can be approximated by the Pomeron exchange, are possible.
}

\begin{document}
\maketitle
\flushbottom

\section{Introduction}
\label{sec:intro}

~~~~Investigating cross sections of high energy hadron-hadron scattering, which reflects the internal structure of hadrons, is one of the most important topics in hadron physics. Forward proton-proton ($\rm pp$) and proton-antiproton ($\rm p \bar p$) scattering at high energies lies mostly outside the regime where the perturbative technique of QCD is applicable. Differential and total cross sections of the forward scattering cannot be rigorously calculated by perturbative or lattice QCD methods.

Before the establishment of QCD, hadronic scattering cross sections had been extensively studied in Regge theory, and in the so-called "classical Regge regime" with the Mandelstam variables $s$ and $t$ satisfying the condition $s \gg |t|$ and $|t|$ of the order or smaller than the QCD scale $\Lambda^2_{\rm QCD}$, the scattering amplitude $\mathcal A (s,t)$ behaves as
\begin{equation}
\mathcal A (s,t) \sim s^{\alpha(t)}.
\label{eq:scattering_amplitude}
\end{equation}
Here $\alpha(t)$ is a linear function of $t$, which is known as the Regge trajectory~\cite{Veneziano:1968yb}:
\begin{equation}
\alpha(t) = \alpha(0) + \alpha^{\prime}t,
\label{eq:regge_trajectory}
\end{equation}
where $\alpha(0)$ and $\alpha^{\prime}$ are constants.
This linear behaviour is supported by the experimental findings that mesons can be classified into groups by a linear relation,
\begin{equation}
J= \alpha(0) + \alpha^{\prime} m^2,
\label{eq:linear_relation}
\end{equation}
where $J$ and $m$ represent the meson spins and masses respectively.
This relation implies that the hadron-hadron scattering can be interpreted as an infinite sum of exchange of mesons lying on the Regge trajectories.

In the recent years, the TOTEM collaboration at the LHC has published their results of the forward $\rm pp$ cross section measurements at centre of mass energies $\sqrt{s}=2.76$~TeV~\cite{Antchev:2018rec}, 7~TeV~\cite{Antchev:2013gaa}, 8~TeV~\cite{Antchev:2016vpy} and 13~TeV~\cite{Antchev:2017dia,Antchev:2017yns,Antchev:2018edk}.
Since 1960s, there have been a large number of  $\rm pp$ and $\rm p \bar p$ cross section measurements.
The centre of mass energy spans from fixed target experiments at $\sqrt{s} \sim 10$~GeV to the LHC energies of $\sqrt{s}
\sim 10$~TeV.
Focusing on the TeV scale, except the TOTEM, there are
earlier data at $\sqrt{s}=546$~GeV measured by the UA4 collaboration~\cite{Bozzo:1984ri,Bozzo:1985th} and the CDF collaboration~\cite{Abe:1993xx},  and at $\sqrt{s}=1.8$~TeV measured by the E710 collaboration~\cite{Amos:1988ng,Amos:1989at,Amos:1991bp} and the CDF collaboration~\cite{Abe:1993xx}.
These measurements have accumulated a rich set of data, upon which various theoretical and phenomenological models have been developed~\cite{Froissart:1961ux,Gauron:1992zc,Block:1994pp,Cudell:2001pn,Cudell:2002xe}.
However, it is found that these models are not enough to well describe all the new TOTEM data from $\sqrt{s}=$ 2.76 to 13~TeV simultaneously~\cite{Antchev:2017yns}.
Hence, more advanced models are certainly needed to explain the data at the LHC energies.

By virtue of the optical theorem:
\begin{equation}
\sigma _{tot} = \frac{1}{s} \mathrm{Im} \mathcal{A} (s,t=0),
\label{eq:optical_theorem}
\end{equation}
from eq.~\eqref{eq:scattering_amplitude} one finds that the total cross sections in the Regge limit behave as $s^{\alpha(0)-1}$.
Experimental data shows that the total cross section of the high energy $\rm pp$ scattering grows with $s$, but all the known meson trajectories called Reggeons have their intercepts $\alpha(0) < 0.6$, which cannot account for the growing behaviour.
In order to explain this behaviour, a new kind of trajectory called Pomeron was introduced~\cite{Collins:1974en}.
The leading Pomeron trajectory has its intercept $\alpha(0) \approx 1$ but slightly greater than 1, which is called
soft Pomeron intercept and can account for the growing behaviour.
Many works have been done to fit the total cross section data by combining the
contributions from the Reggeon and the Pomeron exchange~\cite{Donnachie:1992ny,Cudell:1999tx}.
One may interpret the Pomeron exchange as the multi-gluon exchange, however, it is extremely difficult to derive the soft Pomeron intercept from QCD itself due to its nonperturbative nature.
Therefore, the description by the Pomeron exchange is still useful, and further investigations of the Pomeron dynamics are certainly needed.

The birth of the string theory is closely related to Regge theory.
The Veneziano amplitude which describes the scattering of open strings was originally proposed to model the pion-pion scattering.
The string amplitude can explicitly reproduce the Regge behaviour of the differential cross section.
These are the early interplay between string theory and strong interaction.
Later on QCD was accepted as the fundamental theory of strong interaction and string theory was no longer interpreted as a theory of strong interaction.
However, analytically solving nonperturbative QCD is a formidable task and various hadronic properties cannot be directly derived from QCD.
This dilemma led us to string theory again, but in a different way.
Recently, the anti-de Sitter/conformal field theory  (AdS/CFT) correspondence~\cite{Maldacena:1997re,Gubser:1998bc,Witten:1998qj}, which relates a 4-dimensional conformal field theory to a gravitational theory in the higher dimensional AdS space, provides us a hopeful way to investigate strongly coupled  quantum field theories.
The information of those strongly coupled theories can be obtained by studying their dual theories in the AdS space.
The application of the AdS/CFT correspondence to QCD has gathered a lot of theoretical interests, and many holographic QCD models have been proposed so far.
There are basically two approaches: the top-down approach~\cite{Kruczenski:2003be,Kruczenski:2003uq,Sakai:2004cn,Sakai:2005yt} starting from string theory and the bottom-up approach~\cite{Son:2003et,Erlich:2005qh,DaRold:2005mxj} starting from the known QCD phenomenology to construct the dual theory in the higher dimensional AdS space.
Most of these works are dedicated to hadron properties such as masses, widths, decay constants~\cite{Sakai:2004cn,Sakai:2005yt,Hirn:2005nr,Karch:2002sh,Erlich:2005qh,DaRold:2005mxj,deTeramond:2005su,Brodsky:2006uqa,Karch:2006pv,Abidin:2008ku,Abidin:2008hn,Abidin:2009hr}.
More systematic graviton-dilaton-meson framework has been developed in refs.~\cite{Csaki:2006ji,Gursoy:2007cb,Gursoy:2007er,Li:2013oda}.
A comprehensive review is provided by the authors of ref.~\cite{Brodsky:2014yha}.
There are also many applications of the AdS/CFT correspondence to high energy scattering phenomena~\cite{Polchinski:2001tt,Polchinski:2002jw,Brower:2006ea,Hatta:2007he,Pire:2008zf,Marquet:2010sf,Watanabe:2012uc,Watanabe:2013spa,Watanabe:2015mia,Watanabe:2018owy}, aiming at a better understanding of the internal structure of hadrons.

Based on the string theory, a holographic QCD model has been proposed to describe the elastic $\rm pp$ and $\rm p \bar p$ scattering in the Regge regime~\cite{Domokos:2009hm,Domokos:2010ma,Hu:2017iix}.
The main feature of this model is that the $\rm pp$ scattering is described by the Pomeron exchange, which is realized by combining the Reggeized spin-2 particle propagator and the proton gravitational form factor to dictate the proton-Pomeron vertex.
This model has also been applied to $\eta$ and $\eta^\prime$ central production processes~\cite{Anderson:2014jia,Anderson:2016zon}.
The gravitational form factor used in these works is approximated by the dipole form factor $A(t)=(1-t/M_d^2)^{-2}$, where the dipole mass $M_d$ is the one of the four model parameters. The dipole form factor was originally proposed to fit the elastic electron-proton scattering data at large angles~\cite{Albrecht:1965ki}.
Many studies on the electromagnetic form factors have been done so far, but on the other hand, the gravitational form factors (are also called stress tensor or energy-momentum tensor form factors) are less studied. The gravitational form factors are important physical quantities to reveal the hadron nature.
For instance, one of them measures the total angular momentum carried by partons, which is related to the generalized parton distribution (GPD) functions through sum rules~\cite{Ji:1996ek}.

In this work, the treatment of the proton-Pomeron coupling in previous studies is improved, and the comparison between the model calculations and the newly published TOTEM data for both the differential and total cross sections is explicitly demonstrated.
In our model setup, we use the bottom-up AdS/QCD model to obtain the proton gravitational form factor which can be extracted from the proton-Pomeron(graviton)-proton three-point function.
We consider two versions of the bottom-up AdS/QCD model, the hard-wall model and the soft-wall model.
In the former one, the AdS geometry is sharply cut off  in the infrared (IR) region to introduce the QCD scale, while the geometry is smoothly cut off by utilizing the background dilaton field in the latter one.
The form factors we adopt originally have a few parameters, but they are uniquely fixed by the hadron properties, e.g., the proton mass.
Hence, there is no adjustable parameter in the expression of the proton-Pomeron coupling in our model setup, and we have only three parameters in total which are all from the Reggeized spin-2 particle propagator.
Those parameters are fitted to the available data in the considered high energy regime, taking into account the differential and total cross sections simultaneously.

This paper is organized as follows.
In section~\ref{sec:model} we introduce the model to describe the high energy $\rm pp$ scattering in the Regge regime.
The amplitude of the $2^{++}$ glueball exchange is presented, and then the propagator is Reggeized to describe the Pomeron exchange.
We introduce the hard-wall and soft-wall AdS/QCD model in section~\ref{sec:form_factor}, from which the required gravitational form factor of a proton is obtained.
The kinematic range considered in this study is explained in detail, and our numerical results for the differential and total $\rm pp$ cross sections are presented in section~\ref{sec:fitting}. At last in section~\ref{sec:summary} we give a summary and discuss the implications of the results.

\section{Holographic description of proton-proton scattering in the Regge regime}
\label{sec:model}

The formalism which describes the $\rm pp$ scattering in the Regge regime with the Pomeron exchange has been
developed in ref.~\cite{Domokos:2009hm}.
In the model, the scattering amplitude is obtained by
combining the Pomeron propagator and the gravitational form factor of the proton.
There are two important ingredients in this framework:
(1) The coupling of the Pomeron to the proton is dictated by the vertex of the lowest state on the leading Pomeron trajectory, which is assumed to be the $2^{++}$ glueball.
The $2^{++}$ glueball state has been calculated in the holographic QCD models~\cite{Brower:2000rp} and lattice QCD~\cite{Meyer:2004jc}, and its propagator has been given in ref.~\cite{Yamada:1982dx}.
(2) The $2^{++}$ glueball propagator is then Reggeized to take into account all the states on the Pomeron trajectory.
This is done by exploring the connection between Regge theory and string theory.
In the Regge regime, string amplitude has similar pole and residue structure as in Regge theory.
Closed string amplitude is used to model the exchange of a
trajectory of glueball states, i.e., the Pomeron.

Firstly we explain the derivation of the amplitude of  the $2^{++}$ glueball exchange.
The $2^{++}$ glueball field is expressed as a second-rank symmetric traceless tensor $h_{\mu\nu}$, which is assumed to be coupled predominantly to the QCD energy-momentum tensor $T_{\mu\nu}$:
\begin{equation}\label{eq:glueball}
S  = \lambda \int d^4 x h_{\mu \nu} T^{\mu \nu}.
\end{equation}
Then the proton-glueball-proton vertex can be extracted from the matrix element of the energy-momentum tensor $T_{\mu\nu}$ between the proton states,
\begin{equation}\label{eq:element}
\langle
p', s'| T_{\mu \nu} (0) | p, s  %
\rangle.
\end{equation}
Considering the symmetry and conservation of $T_{\mu \nu}$, eq.~\eqref{eq:element} can be generally expressed in terms of three form factors~\cite{Pagels:1966zza},
\begin{equation} \label{eq:threeff}
\begin{split}
\langle p', s'| T_{\mu \nu} (0) | p, s \rangle = %
\bar u(p', s') \biggl[ &A (t) \frac{ \gamma_\mu P_\nu + \gamma_\nu P_\mu }{2}  \\  %
+ & B(t) \frac{i (P_\mu \sigma_{\nu \rho} + P_\nu \sigma_{\mu \rho}) k^\rho}{4m_p} \\  %
 +  &C(t) \frac{(k_\mu k_\nu - \eta_{\mu \nu } k^2)}{m_p}   \biggr] u(p, s), %
\end{split}
\end{equation}
where
$k = p' - p$, %
$t = k^2$ %
and %
$P = (p + p') / 2$.
At zero momentum transfer, there are two constraints on the form factors:
$A(0)=1$ and $B(0)=0$, due to the fact that the proton has spin $1/2$ and mass $m_p$.
We will see below that we only need to consider the form factor $A(t)$ in this model.
In section~\ref{sec:form_factor} we will explain the calculation of $A(t)$ within the bottom-up AdS/QCD model.

\begin{figure}[t]
	\begin{center}
		\includegraphics[width=0.35\textwidth]{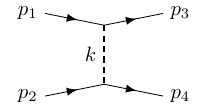}
		\caption{\label{fig:diagram}
The Feynman diagram of the $\rm pp$ scattering by the exchange of a glueball in the t-channel.
$p_1$ and $p_2$ are the four-momenta of the incoming protons, and $p_3$ and $p_4$ are four-momenta of the outgoing protons.
$k$ is the four-momentum of the exchanged glueball.
}
	\end{center}
\end{figure}

The following scattering process is considered: $\rm pp$ (or
$\rm p \bar p$) are scattered by the exchange of a massive, spin-2 glueball.
Only $t$-channel is needed to be considered because it is the dominant channel in the Regge regime.
The Feynman diagram describing the process is shown in figure~\ref{fig:diagram}.

The massive spin-2 glueball propagator can be written as~\cite{Yamada:1982dx}
\begin{equation} \label{eq:propagator}
\frac{d_{\alpha\beta\gamma\delta}(k)} %
{k^2 - m_g^2},
\end{equation}
where $\alpha$ and $\beta$ are Lorentz indices contracted at one side, and $\gamma$ and $\delta$ are the Lorentz indices contracted at the other side.
$m_g$ is the mass of the glueball and $d_{\alpha \beta \gamma \delta}$ can be explicitly expressed as
\begin{equation}
\begin{split}
d_{\alpha \beta \gamma \delta} &=
\frac{1}{2} (\eta_{\alpha \gamma} \eta_{\beta \delta}
+ \eta_{\alpha \delta} \eta_{\beta \gamma})
- \frac{1}{2m_g^2}(k_{\alpha} k_{\delta} \eta_{\beta \gamma}
+ k_{\alpha} k_{\gamma}\eta_{\beta \delta} +
k_{\beta} k_{\delta} \eta_{\alpha \gamma}
+ k_{\beta} k_{\gamma} \eta_{\alpha \delta}) \\ %
& + \frac{1}{24}\left[\left( \frac{k^2}{m_g^2} \right)^2 -
3\left(\frac{k^2}{m_g^2} \right)
- 6\right] \eta_{\alpha \beta} \eta_{\gamma \delta}
- \frac{k^2 - 3 m_g^2}{6 m_g^4}(k_{\alpha}k_{\beta}\eta_{\gamma \delta}
+ k_{\gamma} k_{\delta} \eta_{\alpha \beta}) \\ %
&+ \frac{2k_{\alpha}k_{\beta}k_{\gamma}k_{\delta}}{3 m_g^4}.
\end{split}
\end{equation}
By combing the form factors in eq.~\eqref{eq:threeff} and the propagator in eq.~\eqref{eq:propagator}, the amplitude can be written down as
\begin{equation}\label{eq:amplitude}
\begin{split}
\mathcal M_g &= \frac{\lambda^2d_{\alpha \beta \gamma \delta}}{4 (t - m_g^2)} \\ %
& \times \Big[A(t)(\bar{u}_{1} \gamma^{\alpha} u_{3})(p_1 + p_3)^{\beta} + \frac{i B(t)}{2 m_p}(p_1 + p_3)^{\beta}k_{\rho}(\bar{u}_{1}\sigma^{\alpha \rho}u_{3}) + \frac{C(t)}{m_p}(\bar{u}_1 u_{3}) (k^{\alpha} k^{\beta} - \eta^{\alpha \beta}t)\Big] \\ %
&\times \Big[A(t) (\bar{u}_{2}\gamma^{\gamma} u_{4})(p_2 + p_4)^{\delta} + \frac{i B(t)}{2 m_p} (p_2 + p_4)^{\delta} k_{\lambda}(\bar{u}_{2}\sigma^{\gamma \lambda}u_{4}) + \frac{C(t)}{m_p}(\bar{u}_2 u_{4}) (k^{\gamma}k^{\delta} - \eta^{\gamma \delta} t)\Big].
\end{split}
\end{equation}
Using the condition $s \gg |t|$, it can be seen that the contributions from the $C(t)$ related terms are suppressed by the factor $|t|/s$, and the contributions from the $B(t)$ related terms are negligible compared to the contributions from the $A(t)$ related terms.
So in the Regge regime, eq.~\eqref{eq:amplitude} is greatly reduced and only the terms containing the form factor $A(t)$ need to be considered. After taking these approximations, eq.~\eqref{eq:amplitude} can be expanded and then rearranged into
\begin{equation}\label{eq:finalamp}
\mathcal M_g= \frac{\lambda^2}{8 (t - m_g^2)} \left[ 2 s A^2 (t) (\bar{u}_1 \gamma^{\alpha} u_3) (\bar{u}_2 \gamma_{\alpha} u_4) + 4 A^2 (t) p_2^{\alpha} p_1^{\beta}(\bar{u}_1 \gamma_{\alpha} u_3) (\bar{u}_2 \gamma_{\beta} u_4) \right].
\end{equation}
Since the differential cross section is given by
\begin{equation}\label{eq:Reggedifferential}
\frac{d \sigma}{dt} = \frac{1}{16 \pi s^2} | \mathcal M_g (s, t)|^2,
\end{equation}
the expression for the considered process can be derived by taking the modulus and the spin averaged sum of eq.~\eqref{eq:finalamp}.
We obtain
\begin{equation}\label{eq:dif}
\frac{d \sigma}{dt} = \frac{\lambda^4 s^2 A^4 (t)}{16 \pi (t - m_g^2)^2}.
\end{equation}
This differential cross section only represents the exchange of the lightest state, i.e., the $2^{++}$ glueball.
In order to include the higher spin states on the Pomeron trajectory, a procedure named Reggeizing is employed to obtain the Pomeron propagator from the string amplitude.

The simplest closed string scattering is the scattering of four closed string tachyons.
The scattering amplitude can be written in the form
\begin{equation}
\mathcal M_c = %
\frac{\Gamma[-a_c (t)] \Gamma[-a_c (u)] \Gamma[-a_c (s)]}{\Gamma[-a_c (t) - a_c(s)]\Gamma[-a_c (t) - a_c (u)] \Gamma[-a_c (u) - a_c (s)]} K_c (p_1, p_2, \dots),
\end{equation}
where $s$, $t$ and $u$ are Mandelstam variables and $K_c$ is a kinematic factor with no poles, which depends on the momenta and polarizations of the scattered particles.
$a_c(x)=a_c(0) + a_c'x$ is a linear function related to the spectrum of the closed strings.
The parameters $a_c (0)$ and $a^{\prime}_c $ are related to the Pomeron trajectory parameters through
\begin{equation}
2 a_c (0) + 2 = \alpha_c (0), \ \ \ 2 a_c' = \alpha_c'.
\end{equation}
In the Regge regime, the amplitude can be written as
\begin{equation} \label{eq:poles}
\mathcal M_c^{Reg} = e^{- i \pi a_c (t)} (a'_c s)^{2 a_c (t)} \frac{\Gamma[-a_c (t)]}{\Gamma[a_c (t) - \chi]} K_c (p_1, p_2, \dots),
\end{equation}
where $ \chi $ is defined by $\chi  =  a_c (s) + a_c (t) + a_c (u) = 4 a'_c m^2 + 3 a_c (0)$, with $m$ the mass of the scattered particles.
From eq.~\eqref{eq:poles} it can be seen that the amplitude has poles at $a_c (t) = n$ for $n = 0, 1, 2, \cdots$ with residue $\sim s^{2 n}$.
This is assumed to be corresponding to the $t$-channel exchange of a trajectory with $J = 2 n = 2, 4, 6 \cdots$.

In the Regge regime the amplitude for the exchange of the lowest state on the trajectory is given by
\begin{equation}\label{eq:1st}
\mathcal{M}_{\rm 1st} \approx \frac{- s^{2} f(\epsilon_i)} %
{a_c' \Gamma[-\chi] (t - m_g^2)},
\end{equation}
where $f (\epsilon_i)$ is some unknown function of the polarizations of the scattered particles. The full amplitude is given by
\begin{equation}\label{eq:full}
\frac{\Gamma \left[1 - \frac{\alpha_c (t)}{2}\right]}{\Gamma \left[ \frac{\alpha_c (t)}{2} -1 - \chi \right]} e^{- i \pi \alpha_c (t) / 2} \left(\frac{\alpha'_c s}{2} \right)^{\alpha_c (t) - 2} s^{2}f (\epsilon_i).
\end{equation}
There is a factor between eq.~\eqref{eq:full} and eq.~\eqref{eq:1st}:
\begin{equation}
\frac{- \alpha_c'}{2}\frac{\Gamma [-\chi] \Gamma \left[1 - %
\frac{\alpha_c (t)}{2}\right]}{\Gamma \left[\frac{\alpha_c (t)}{2} -1 - \chi\right]} e^{- i \pi \alpha_c (t) / 2}\left(\frac{\alpha'_c s}{2} \right)^{\alpha_c (t) - 2}.
\end{equation}
The differential cross section of the Pomeron exchange can be obtained by replacing the glueball exchange factor $\frac{1}{t - m_g^2}$ in eq.~\eqref{eq:dif} with the above factor,
\begin{equation}\label{eq:diff}
\frac{d \sigma}{dt} = \frac{\lambda^4 A^4 (t) \Gamma^2[- \chi] \Gamma^2 \left[1 - %
\frac{\alpha_c (t)}{2}\right]}{16 \pi \Gamma^2 \left[\frac{\alpha_c (t)}{2} - 1 - \chi \right]} \left(\frac{\alpha'_c s}{2}\right)^{2 \alpha_c (t) - 2},
\end{equation}
which corresponds to the Pomeron exchange amplitude,
\begin{equation}
\mathcal{M} (s, t) = s \lambda^2 A^2 (t) e^{- i \pi \alpha_c (t) / 2} \Biggl( %
\frac{\Gamma[-\chi]\Gamma\left[1 - %
\frac{\alpha_c (t)}{2}\right]}{\Gamma\left[\frac{\alpha_c (t)}{2} -1 - \chi \right]} \Biggr) \left(\frac{\alpha'_c s}{2} \right)^{\alpha_c (t) - 1}.
\end{equation}
The total cross section can be obtained by applying the optical theorem to eq.~\eqref{eq:full}:
\begin{equation}\label{eq:total}
\sigma_{tot} = \frac{1}{s} \ {\rm Im} \ \mathcal{M} (s, 0) = %
\frac{\pi \lambda^2 \Gamma [- \chi]}{\Gamma %
	\left[\frac{\alpha_c (0)}{2} \right]\Gamma %
    \left[\frac{\alpha_c (0)}{2} - 1 - \chi \right]} %
\left(\frac{\alpha_c' s}{2} \right)^{\alpha_c (0) - 1}.
\end{equation}
%

\section{Proton gravitational form factor from the bottom-up AdS/QCD model}
\label{sec:form_factor}

The gravitational form factor of a nucleon has been calculated in a bottom-up AdS/QCD model~\cite{Abidin:2009hr}, in which a fermion field couples to a vector field in the 5-dimensional AdS space, and the gravitational form factor is calculated by perturbing the metric of the static AdS solution. Here we briefly review the previous study, in which the model action is given by
\begin{equation}
S_F = %
\int d^5 x \sqrt{g} e^{- \Phi (z)} \bigg( \frac{i}{2} \bar{\Psi} e^N_A \Gamma^A D_N \Psi %
- \frac{i}{2} (D_N \Psi)^\dagger \Gamma^0 e^N_A \Gamma^A \Psi - (M + \Phi (z)) \bar{\Psi}\Psi \bigg),
\end{equation}
where $e^N_A = z \delta^N_A$, $D_N = \partial_N + \frac{1}{8} \omega_{NAB}[\Gamma^A, \Gamma^B] - i V_N$ is the covariant derivative and $M$ is the mass of the bulk spinor.
The Dirac gamma matrices are defined by the anti-commutation relation $\{ \Gamma^A,\Gamma^B \} = 2 \eta^{AB}$.
The soft-wall model is implemented by adding $\Phi (z) = \kappa^2 z^2$ to the mass term.
From the action one can derive the equation of motion of the Dirac field,
\begin{equation}
\biggl[ i e^N_A \Gamma^A D_N - \frac{i}{2} (\partial_N \Phi) e^N_A \Gamma^A %
 - (M + \Phi (z))\biggr] \Psi = 0.
\end{equation}
Evaluating the action on the solution gives
\begin{equation}
S_F [\Psi_{cl}] = %
\int d^4 x \frac{- 1}{2 z^4} e^{- \kappa^2 z^2} %
\left(\bar \Psi_L \Psi_R - \bar \Psi_R \Psi_L \right) %
\Big|_\epsilon^{z_{IR}},
\end{equation}
where $\Psi_{R, L} = (1 / 2) (1 \pm \gamma^5) \Psi$ and $\epsilon$ represents the ultraviolet (UV) boundary.
In the hard-wall model ($\kappa = 0$) the IR cutoff is at $z_{IR}=z_0$,  and in the soft-wall model $z$ extends to infinity, i.e., $z_{IR} = \infty$. In order to preserve the  $O (5, 1)$ isometry group of the original action, an extra term is added in the UV boundary,
\begin{equation}
\frac{1}{2} \int d^4 x \sqrt{-g^{(4)}} %
\left( \bar{\Psi}_L %
\Psi_R + \bar{\Psi}_R %
\Psi_L \right)_{\varepsilon},
\end{equation}
which corresponds to the action,
\begin{equation}
S_F = %
\int d^4 x \left( \frac{1}{z^4} \bar{\Psi}_L %
\Psi_R \right)_{\varepsilon}.
\end{equation}
The Dirac fields in the momentum space can be written as $\Psi_{R, L} (p, z) = z^\Delta f_{R, L} (p, z) \Psi^0 (p)_{R, L}$, in which $\Psi^0_{R, L}$ are the 4-dimensional boundary fields and $f_{R, L}$ are the profile functions or the bulk-to-boundary propagators.
$\Psi^0_L (p)$ is chosen as the independent source field which corresponds to the spin-$\frac{1}{2}$ baryon operator $\mathcal{O}_R$ in the 5-dimensional field theory, and $ \Delta$ is chosen such that the equation of motion allows $f_L (p, \varepsilon ) = 1$.

It is known that the left handed and the right handed components of the spin-$\frac{1}{2}$ field operators in the 5-dimensional flat space are related by the Dirac equation, which means that $\Psi^0_L$ and $\Psi^0_R$ are not independent.
There is a relationship $\slashed{p} \Psi^0_R (p) = p \Psi^0_L (p)$ between the left handed and the right handed components.
Dropping the interaction term with the vector field, the equation of motion for the Dirac field can be written as
\begin{eqnarray}\label{fLfR}
\left(\partial_z - \frac{2 + M - \Delta + 2 \Phi}{z} \right) f_R &=&-pf_L, \\ %
\left(\partial_z - \frac{ w - M - \Delta} {z} \right) f_L &=&pf_R.
\end{eqnarray}
In both the soft-wall and the hard-wall models, the equations of motion of the profile functions can be analytically solved.
Besides the gravitational form factor, the electromagnetic form factors were also studied in ref.~\cite{Abidin:2009hr}.
The mass $M$ in the above equations can be fixed by taking into account the large momentum scaling of the obtained electromagnetic form factors, and the resulting value is $M = \frac{3}{2}$.

In addition to the fermion part, there is a kinetic part of the vector field,
\begin{equation}
S_V = \int d^{5} x e^{-\Phi} \sqrt{g} %
{\rm Tr} \left( - \frac{F^2_V}{2 g^2_5} \right),
\end{equation}
where $F^V_{M  N} = \partial_M V_N - \partial_N V_M $.
The transverse part of the vector field can be expressed as $V_\mu (p, z) = V (p, z) V^0_\mu (p)$, and the bulk-to-boundary propagator satisfies $V (p, \varepsilon) = 1$ at the UV boundary.
According to the holographic dictionary, the $V^0_\mu (p)$ is the source for the 4-dimensional current operator $J^V_\mu$. In the $V_z = 0$ gauge, the equation of motion can be obtained~\cite{Grigoryan:2007my}:
\begin{equation} \label{eq:vector}
\left[ \partial_z \left( \frac{e^{-\Phi}}{z} \partial_z \right) %
+ \frac{e^{-\Phi}}{z}p^2\right] %
V (p, z) = 0.
\end{equation}
The solution of the above equation is a normalizable mode with its eigenvalue $p^2 = M^2_n$ which corresponds to the mass of the $n$-th Kaluza-Klein mode of the vector meson~\cite{Erlich:2005qh}.
In the soft-wall model~\cite{Grigoryan:2007my}, the mass eigenvalues are $M^2_n = 4 \kappa^2 (n + 1)$, where $n=0,1,\ldots$.
In the hard-wall model, the eigenvalues are expressed as $M_n = \gamma_{0, n + 1} / z_0$, in which $\gamma_{0, n + 1}$ is the $n + 1$-th zeros of the Bessel function $J_0$.

The stress tensor matrix element for spin-$\frac{1}{2}$ particles has been generally written in terms of the three form factors in eq.~\eqref{eq:threeff}, which can be extracted from the following 3-point function,
\begin{equation} \label{eq:3-point}
\left< 0 \big| \mathcal{T} \mathcal{O}^i_R (x) %
T^{\mu \nu} (y) \bar{ \mathcal{O}}^j_R (w) \big| 0 \right>.
\end{equation}

The 5-dimensional AdS space metric is given by
\begin{equation}
ds^2 = g_{MN} dx^M dx^N = %
\frac{1}{z^2} \left(\eta_{\mu \nu} dx^\mu dx^\nu - dz^2 \right).
\end{equation}
The energy-momentum tensor operator in the 4-dimensional strongly coupled theory corresponds to the metric perturbation in the 5-dimensional gravity theory.
In order to calculate the gravitational form factors, we consider the gravity-dilaton-tachyon action~\cite{Batell:2008zm,Batell:2008me}, and in which the metric of the AdS space is perturbed from its static solution according to $\eta_{\mu \nu} \to \eta_{\mu \nu} + h_{\mu \nu} $.
The action in the second order perturbation can be written as
\begin{equation}
S_{GR} = %
- \int d^5 x \frac{e^{- \kappa^2 z^2}}{4 z^3} %
\left( h_{\mu \nu, z}{h^{\mu \nu}}_{,z} %
+ h_{\mu \nu} \Box h^{\mu \nu} \right),
\label{eq:action_gr}
\end{equation}
where $h_{\mu \nu}$ satisfies the transverse-traceless gauge conditions, $\partial^\mu h_{\mu \nu} = 0$ and $h^\mu_\mu = 0$.
From the action one can find that the following Einstein equation is satisfied by the profile function of the metric perturbation:
\begin{equation}
\left[ \partial_z \left(\frac{e^{- \kappa^2 z^2}}{z^3} \partial_z \right) %
+ \frac{e^{- \kappa^2 z^2}}{z^3} p^2 \right] h(p, z) = 0.
\label{eq:ein_eq}
\end{equation}
This equation can be solved in both the soft-wall and the hard-wall models.
In the former, the solution is given with the Kummer function by
\begin{equation}
\begin{split}
H (Q, z) &= \Gamma(a + 2) U (a, - 1 ; \xi ) \\ %
&=a (a + 1) \int_0^1 dx x^{a - 1} (1 - x) %
\exp\left(\frac{ - \xi x}{1 - x} \right),
\end{split}
\label{eq:b-to-b_propagator}
\end{equation}
where $H (Q, z)\equiv h (q^2 = - Q^2, z)$, $a = Q^2 / (4 \kappa^2)$ and $\xi = \kappa^2 z^2$.
The constant $\kappa$ is determined by the proton and the $\rho$-meson masses.
The best fit can be obtained as $\kappa = 0.350$~GeV, which gives the proton mass $0.990$~GeV and the $\rho$-meson mass $0.700$~GeV.
The solution in the soft-wall model satisfies $H (p, \varepsilon) = 1$ and vanishes at infinity. For the hard-wall model, the solution is given with the Bessel functions by
\begin{equation}
H (Q, z) = \frac{(Qz)^2}{2} %
\left( \frac{K_1 (Qz_0)}{I_1 (Qz_0)} I_2 (Qz) + K_2 (Qz) \right).
\end{equation}
The value of $z_0$ that one can obtain with the proton mass is $z_0 = (0.245$~GeV$)^{-1}$.

\begin{figure}[t]
	\begin{center}
		\includegraphics[width=0.5\textwidth]{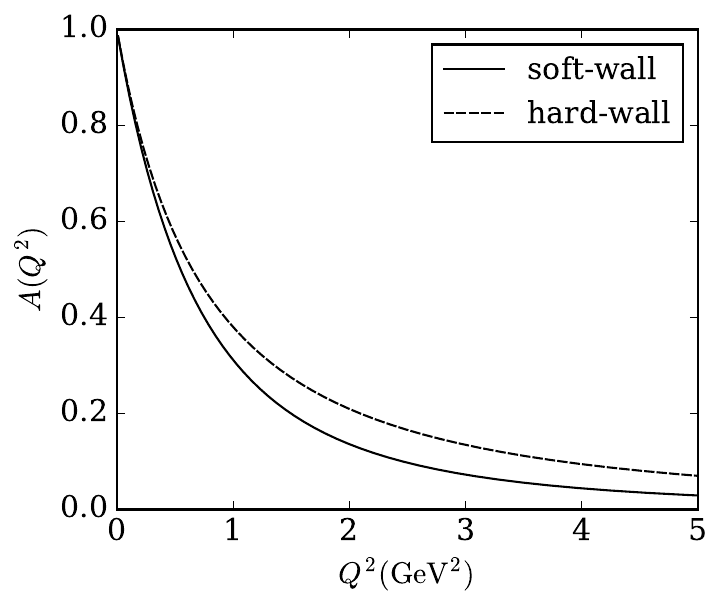}
		\caption{\label{fig:form_factors}
		The gravitational form factor of the proton from a bottom-up AdS/QCD model~\cite{Abidin:2009hr}.
		The solid and dashed lines are from the soft-wall and the hard-wall models, respectively.
		}
	\end{center}
\end{figure}

Combining the solutions for the Dirac fields $\Psi_{L, R} (z)$ and the solution for the metric perturbation, the 3-point function eq.~\eqref{eq:3-point} can be calculated and the gravitational form factor $A (t)$ takes the form of
\begin{equation}
A (Q) = %
\int dz \frac{ e^{- \kappa^2 z^2}}{2 z^{2 M}} H (Q, z) %
\left(\Psi_L^2 (z) + \Psi_R^2 (z) \right).
\end{equation}
The $Q^2$ dependencies of the proton gravitational form factors obtained from both the hard-wall and the soft-wall models are shown in figure~\ref{fig:form_factors}.
These results are comparable with that calculated from a GPD model~\cite{Guidal:2004nd}, while the soft-wall model result is more consistent with the GPD result compared to the hard-wall one.
%

\section{Numerical results}
\label{sec:fitting}

\subsection{Kinematic regions and model parameters}

This model works in the Regge regime, in which we have obtained the expressions of the differential and total cross sections in section~\ref{sec:model}.
When performing the numerical fit to determine the adjustable parameters included in the model, we select experimental data in the kinematic region, where $\sqrt{s} \ge 546$~GeV and $|t| < 0.45$~GeV$^2$, which gives the ratio $|t| / s  < 1.5\times 10^{-6}$.
So the Regge condition is strictly satisfied in this regime.
Since the diffractive minimum (dip) around $|t| = 0.47$~GeV$^2$ has recently been observed in the differential cross section measurement at $\sqrt{s} = 13$~TeV by the TOTEM collaboration~\cite{Antchev:2018edk}, we choose the value mentioned above as the upper cut for $|t|$.

For the smaller $s$ region, the exchange of both the Reggeon and the Pomeron contributes to the total cross sections $\sigma_{tot}$ of the $\rm pp$ and $\rm p \bar p$ scattering.
However, the Reggeon contribution is strongly suppressed at large $s$.
Via the preceding study~\cite{Domokos:2009hm}, it is found that the Reggeon contributions to the $\rm pp$ and $\rm p \bar p$ scattering are less than $1$~\% at $\sqrt{s} = 400$~GeV, and those become about $0.3$~\% at $\sqrt{s} = 1.8$~TeV.
Hence, in this study we consider the experimental data measured in the range, $546$~GeV $\le \sqrt{s} \le 13$~TeV, where it is reasonable to take into account the Pomeron contributions only, neglecting those from the Reggeon.

Theoretically there are two fundamental interactions engaged in the elastic $\rm pp$ scattering: the electromagnetic and the strong interactions.
When the scattering angle is very small (i.e., $|t|$ is very small), the electromagnetic Coulomb interaction plays an important role.
This contribution is of importance at $|t| \approx 0.002$~GeV$^2$ and becomes negligible when $|t| >  0.01$~GeV$^2$~\cite{Amos:1985wx,Bernard:1987vq}.
So we set $|t| = 0.01$~GeV$^2$ as the lower cut.
For the strong interaction part, when $|t|$ is smaller compared to the QCD scale $\Lambda^2_{\rm QCD}$, we stay in the soft region, in which the model of this work is valid.
When $|t|$ becomes larger, we enter the so-called hard Pomeron region, where perturbative QCD starts to be applicable.
There is still no solid theoretical explanation on how and where this soft to hard transition occurs.
The preceding work~\cite{Hu:2017iix} argues that this transition occurs at $t\approx - 0.5$~GeV$^2$.

\subsection{Fitting results}

There are three adjustable parameters in our model: the Pomeron intercept $\alpha_c (0)$, the Pomeron slope $\alpha_c'$ and the proton-glueball-proton coupling constant $\lambda$.
In the analysis presented in ref.~\cite{Domokos:2009hm}, the authors used the dipole form factor, $A (t) = (1 - t / M_d^2)^{- 2}$, with the dipole mass $M_d$ as a free parameter.
In this study we utilize the proton gravitational form factor obtained from the bottom-up AdS/QCD model instead of the dipole one, and this does not bring any additional parameter to the whole model setup.

We fit the differential cross section eq.~\eqref{eq:diff} and the total cross section eq.~\eqref{eq:total} simultaneously to the experimental data, and determine the three parameters $\lbrace \alpha_c (0), \alpha_c', \lambda \rbrace$.
We consider two data sets for the differential cross section.
The first set consists of all the available data in the range, $546$~GeV $\le \sqrt{s} \le 13$~TeV.
We take into account the data measured at $\sqrt{s} = 546$~GeV by the UA4~\cite{Bozzo:1984ri,Bozzo:1985th} and CDF~\cite{Abe:1993xx} collaborations, at $\sqrt{s} = 1.8$~TeV by the E710~\cite{Amos:1988ng,Amos:1989at,Amos:1991bp} and CDF~\cite{Abe:1993xx} collaborations and at $\sqrt{s}=2.76$~TeV~\cite{Antchev:2018rec}, 7~TeV~\cite{Antchev:2013gaa}, 8~TeV~\cite{Antchev:2016vpy} and 13~TeV~\cite{Antchev:2017dia,Antchev:2017yns,Antchev:2018edk} by the TOTEM collaboration.
The second data set consists of the TOTEM data only.
For the total cross section, the TOTEM data~\cite{Antchev:2013gaa,Antchev:2013iaa,Antchev:2013paa,Antchev:2015zza,Antchev:2016vpy,Nemes:2017gut,Antchev:2017dia} and other $\rm pp$~\cite{Baltrusaitis:1984ka,Honda:1992kv,Collaboration:2012wt} and $\rm p \bar p$~\cite{Battiston:1982su,Hodges:1983oba,Bozzo:1984rk,Alner:1986iy,Amos:1991bp,Abe:1993xy,Augier:1994jn,Avila:2002bp} data which  were collected by the Particle Data Group (PDG) in 2010~\cite{Nakamura:2010zzi} are currently available.
In this paper, we present the comparison between the model calculation and all of those data, however, we do not use the data extracted from the cosmic-ray experiments for the fitting, because they have huge uncertainties and are not suitable for the present analysis.
Taking into account the experimental data mentioned above, we perform the log-linear least squares fit, utilizing the MINUIT package in Python~\cite{James:1975dr}.

The results of fitted parameters and $\chi^2 / d.o.f.$, which is calculated using the 8~TeV TOTEM data for the differential cross section, for the two data sets are shown in table~\ref{table:parameters1}
\begin{table}[tb]
\centering
\caption{
The fitting results obtained with the differential cross section data in the range, $546$~GeV $\le \sqrt{s} \le 13$~TeV, and the total cross section data.
} \label{table:parameters1}
\begin{tabular}{c c c  }
\hline
\hline
Parameters \qquad \qquad \qquad & soft-wall \qquad\qquad \qquad&  hard-wall  \\
\hline
$\alpha_c(0)$  \qquad \qquad \qquad   & $1.084\pm 0.005$ \qquad \qquad \qquad
&$1.086\pm 0.003$  \\
$\alpha_c'$  ($\mathrm{GeV}^{-2}$) \qquad \qquad \qquad & $0.368 \pm 0.008$
\qquad \qquad \qquad &$0.377 \pm 0.007$  \\
$\lambda$  ($\mathrm{GeV}^{-1}$) \qquad \qquad \qquad & $9.59\pm 0.44$
\qquad \qquad \qquad  &$9.70\pm 0.32$  \\
$\chi^2/d.o.f.$ \qquad \qquad \qquad & 1.304
\qquad \qquad \qquad  &1.352  \\
\hline
\hline
\end{tabular}
\end{table}
and table~\ref{table:parameters2}
\begin{table}[tb]
\centering
\caption{
The fitting results obtained with the differential cross section data measured by the TOTEM collaboration in the range, $2.76 \le \sqrt{s} \le 13$~TeV, and the total cross section data.
} \label{table:parameters2}
\begin{tabular}{c c c  }
\hline
\hline
Parameters \qquad \qquad \qquad & soft-wall \qquad\qquad \qquad&  hard-wall  \\
\hline
$\alpha_c(0)$  \qquad \qquad \qquad   & $1.086\pm 0.007$ \qquad \qquad \qquad
&$1.087\pm 0.003$  \\
$\alpha_c'$  ($\mathrm{GeV}^{-2}$) \qquad \qquad \qquad & $0.372 \pm 0.011$
\qquad \qquad \qquad &$0.381 \pm 0.010$  \\
$\lambda$  ($\mathrm{GeV}^{-1}$) \qquad \qquad \qquad & $9.60\pm 0.55$
\qquad \qquad \qquad  &$9.75\pm 0.36$  \\
$\chi^2/d.o.f.$ \qquad \qquad \qquad & 1.255
\qquad \qquad \qquad  &1.282  \\
\hline
\hline
\end{tabular}
\end{table}
respectively.
Gravitational form factors obtained from both the soft-wall and the hard-wall models are used in the fits.
We find that the soft-wall model produces the slightly better $\chi^2 / d.o.f.$ values, compared with the hard-wall model results.
Also, it can be seen for both soft-wall and hard-wall model cases that the second data set produces better $\chi^2 / d.o.f.$ values,
comparing with the other data set results.

\begin{figure}[t]
	\begin{center}
		\includegraphics[width=0.99\textwidth]{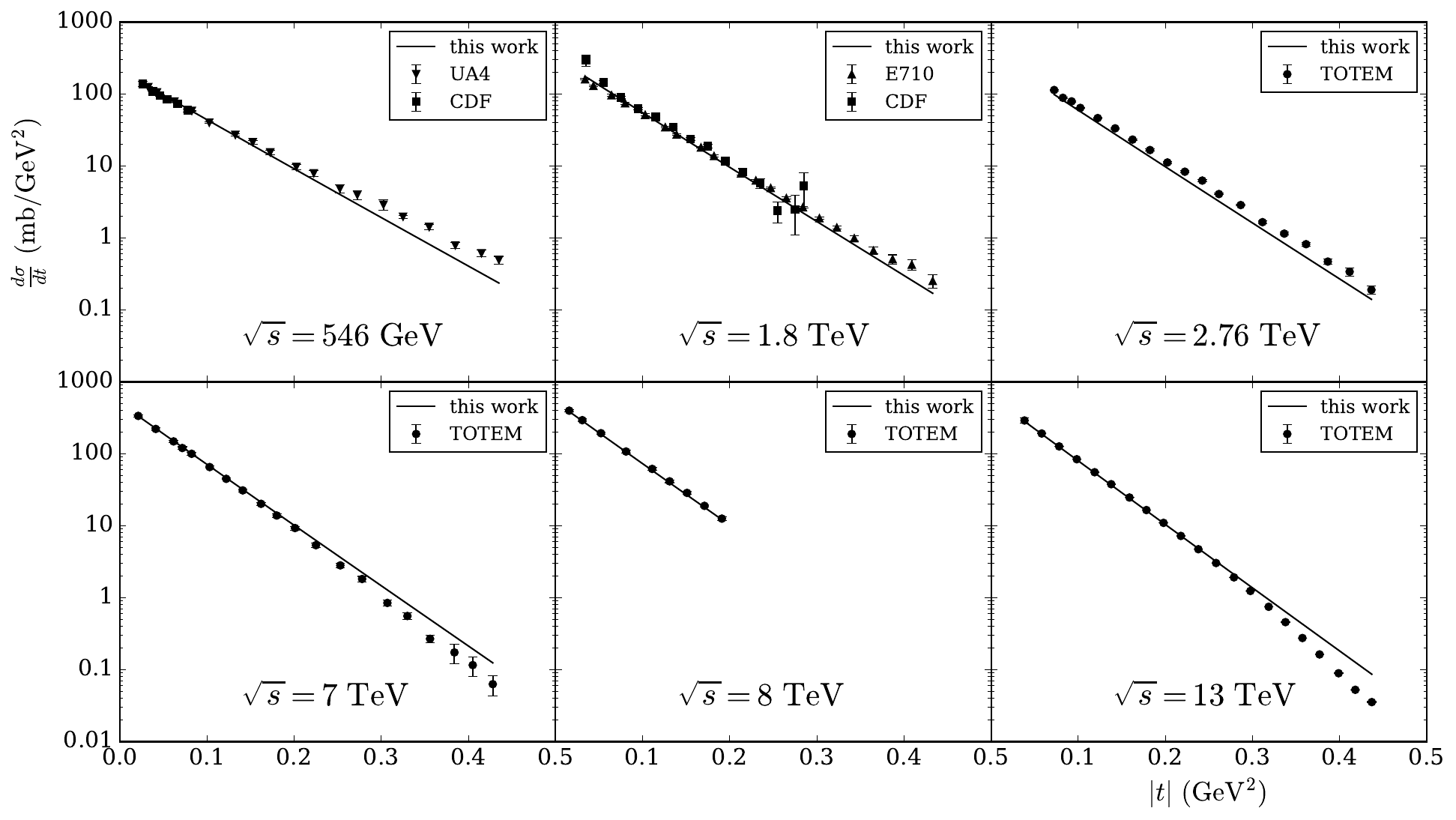}
		\caption{\label{fig:all_soft}
The differential cross section as a function of $|t|$.
The solid lines represent our calculations obtained with the experimental data in the range, $546$~GeV $\le \sqrt{s} \le 13$~TeV.
The data are depicted with their errors.
}
	\end{center}
\end{figure}

\begin{figure}[t]
	\begin{center}
		\includegraphics[width=0.7\textwidth]{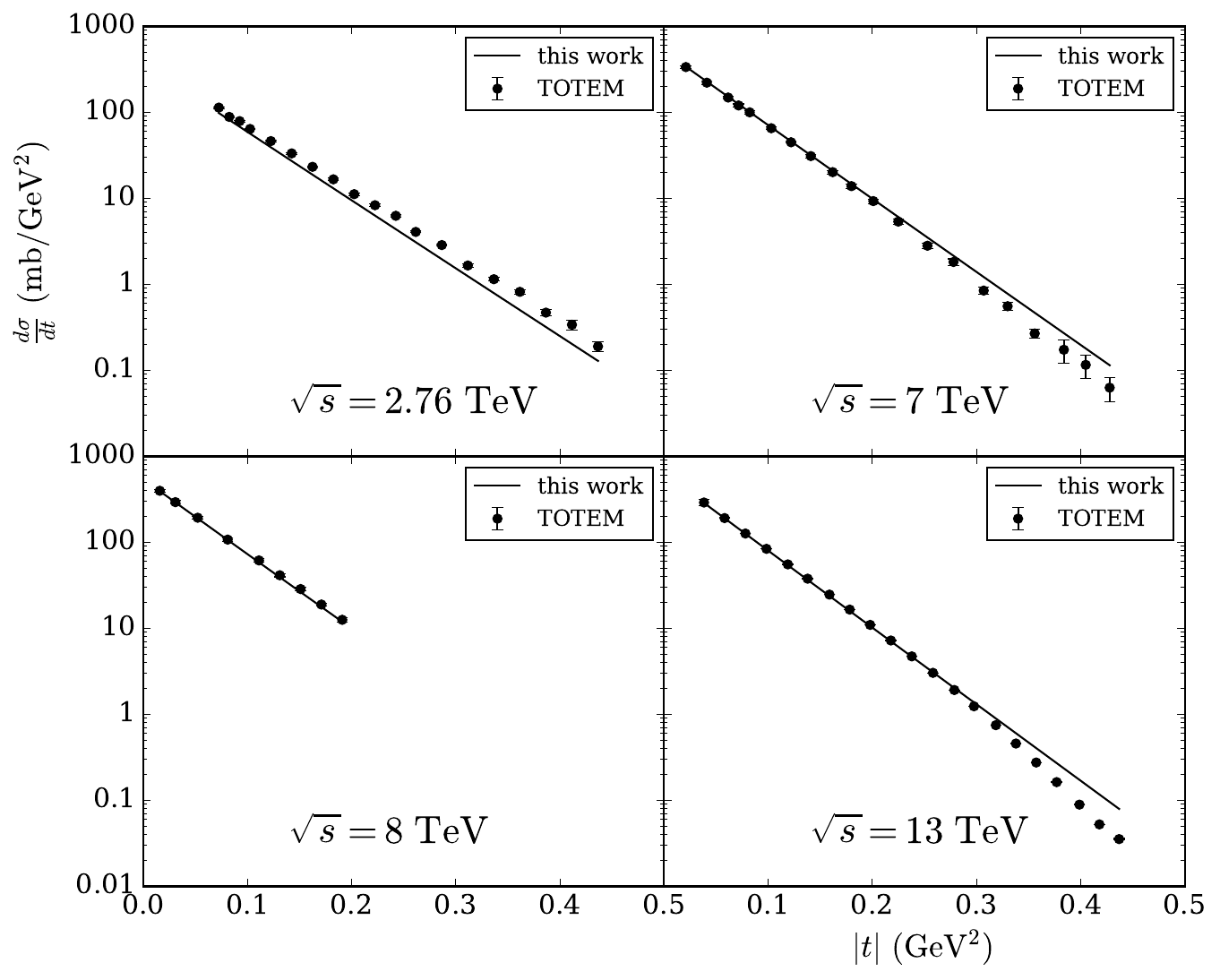}
		\caption{\label{fig:totem_soft}
The differential cross section as a function of $|t|$.
The solid lines represent our calculations obtained with the experimental data measured by the TOTEM collaboration in the range, $2.76 \le \sqrt{s} \le 13$~TeV.
The data are depicted with their errors.
}
	\end{center}
\end{figure}

The results of the differential cross sections for the both data sets are shown in figure~\ref{fig:all_soft}
and figure~\ref{fig:totem_soft}
respectively.
Since the soft-wall and hard-wall model results are quite close to each other, we only present the soft-wall results here.
From these results we find that our calculations are consistent with the data in the considered kinematic regime.
Focusing on the comparisons between our calculations and the TOTEM data, it is seen that the second data set results show better agreements with the data, compared with the other data set results, although this can also be understood from the difference between their $\chi^2 / d.o.f.$ values.

\begin{figure}[t]
	\begin{center}
		\includegraphics[width=0.7\textwidth]{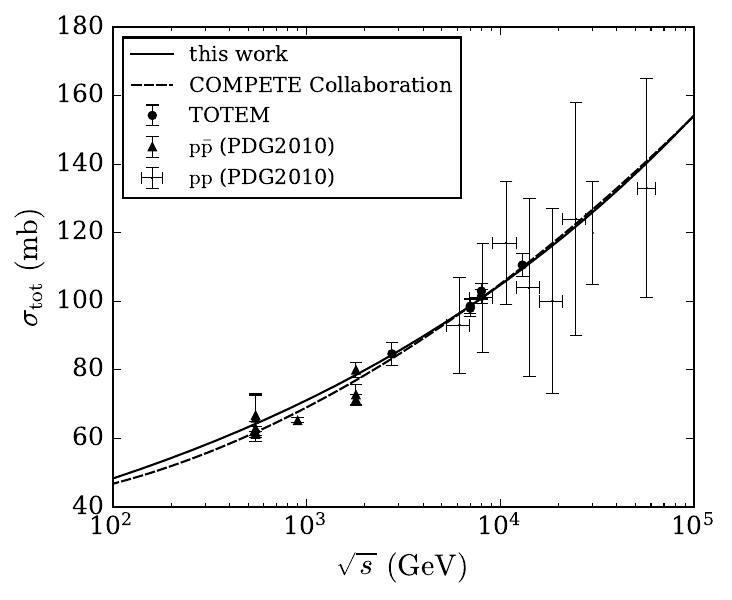}
		\caption{\label{fig:all_soft_total}
The total cross sections of the $\rm pp$ and $\rm p \bar p$ scattering as a function of $\sqrt{s}$.
The solid and dashed curves represent our calculation and the empirical fit performed by the COMPETE collaboration respectively.
The circles and triangles depict the $\rm pp$  data measured by the TOTEM collaboration and the $\rm p \bar p$ data  from earlier experiments respectively.
The $\rm pp$ data from the cosmic-ray experiments are denoted by cross points, but were not used for the fit.
}
	\end{center}
\end{figure}

We show in figure~\ref{fig:all_soft_total} the resulting total cross section of the $\rm pp$ scattering.
Our calculation is obtained with the differential and total cross section data in the range, $546$~GeV $\le \sqrt{s} \le 13$~TeV, and the gravitational form factor calculated within the soft-wall AdS/QCD model.
The $\rm pp$ data extracted from the cosmic-ray experiments are plotted together on the figure, but were not used for the fit.
We find that our calculation agrees with the data in the whole considered kinematic region, where $10^2 < \sqrt{s} < 10^5$~GeV.
Moreover, it can be seen that our result is also quite consistent with the empirical fit performed by the COMPETE collaboration~\cite{Cudell:2002xe}.

From the results presented above, we find that the present model setup can well describe both the differential and total $\rm pp$ cross sections in the considered high energy region.
The obtained value through the numerical fitting of the Pomeron intercept is around $1.086$, which is consistent with that obtained by the authors of ref.~\cite{Domokos:2009hm}, while our Pomeron slope result is around $0.400$~GeV$^2$ and this is larger than their result by $34$~\%.
The resulting value of the coupling constant $\lambda$ is around $9.0$~GeV$^{-1}$, which is slightly larger than their result.
However, they obtained $\lambda = 9.02$~GeV$^{-1}$ via another analysis based on a Skyrme model~\cite{Domokos:2010ma}, and this is quite consistent with ours.

\section{Summary and discussion}
\label{sec:summary}

We have studied the high energy $\rm pp$ scattering cross sections in the framework of holographic QCD, focusing on the Regge regime.
In our model setup, the nonperturbative partonic dynamics is described by the Pomeron exchange, which is realized by applying the Reggeized spin-2 particle propagator and the proton gravitational form factor obtained from the bottom-up AdS/QCD model.
In this study, we have improved the treatment of the gravitational form factor to construct a more consistent model, compared to previous works.
The dipole form factor, which includes the dipole mass as a parameter, was used in ref.~\cite{Domokos:2009hm} to specify the proton-Pomeron coupling.
Hence, there are four adjustable parameters in their model, because the propagator part includes three parameters.
On the other hand, the gravitational form factor we applied in this work does not bring any parameter.
To be precise, it originally includes a few parameters, but those can be uniquely fixed by some hadron properties such as the proton mass.
Therefore, our present model includes only the three parameters in total, which is obviously an advantage.
Also, since the propagator we applied in this study was originally derived based on the closed string amplitude, it can be interpreted as the Reggeized graviton.
So our choice for the gravitational form factor can be easily justified, because it specifies the proton-graviton coupling.

Considering the applicability of our present model and taking into account the recent experimental finding that the diffractive minimum around $|t| = 0.47$~GeV$^2$ was observed in the differential cross section measurement, we decided to focus on the kinematic region, where $546$~GeV $\le \sqrt{s} \le 13$~TeV and $0.01 < |t| < 0.45$~GeV$^2$, and performed the numerical fit.
Currently available most data, except the ones extracted from the cosmic-ray experiments, were used to determine the three adjustable parameters included in the model.
The comparisons between our calculations and the data have been explicitly demonstrated.
We have found that utilizing the soft-wall AdS/QCD model to obtain the gravitational form factor results in slightly better $\chi^2 / d.o.f.$ values, compared to the case of using the hard-wall model.
Moreover, it has been seen that utilizing only the TOTEM data, instead of using all the available data, for the fitting leads to better $\chi^2 / d.o.f.$ values.
A reason for this is that there is a tension between the E710 and CDF data at $\sqrt{s} = 1.8$~TeV, which can lower the fitting quality.
Nonetheless, both the resulting differential and total cross sections are in agreement with the data in the whole considered kinematic region.

The excellent agreement with the data and the empirical fit performed by the COMPETE collaboration for the total cross section shall be emphasized, but this is actually a surprise, because the Froissart bound is not taken into account in the present model.
In general, it is expected that there is some suppression by the Froissart bound and the total cross section grows no faster than $\log^2 s$ in the high $s$ regime.
However, our result implies that the behaviour  $\sigma_{tot} \sim s^{\alpha(0)-1}$ is still a good description in the presently considered kinematic region.
To pin down this, more data are definitely required especially in the higher energy region.

The results we obtained through this study suggest that the present framework has potential to be a useful analytic tool for studies of various high energy scattering processes, in which the involved strong interaction can be approximated by the Pomeron exchange.
Further investigations are certainly needed.
For instance, meson-nucleon or meson-meson scattering can be analyzed by simply replacing the gravitational form factor with the meson's.
Moreover, further applications to more complicated processes, such as the deeply virtual Compton scattering or the electron-ion scattering, shall also be considered.
Some applications are under consideration, and will be reported separately.

\acknowledgments
W.X. is supported by China Three Gorges University under Grant No. 1910103 (20152953), and the NSFC under Grant No. 11647173.
A.W. is supported by Chinese Academy of Sciences (CAS) President's International Fellowship Initiative under Grant No. 2019PM0124 and partially by China Postdoctoral Science Foundation under Grant No. 2018M641473.
M.H. is supported in part by the NSFC under Grant Nos. 11725523, 11735007, 11261130311 (CRC 110 by DFG and NSFC), Chinese Academy of Sciences under Grant No. XDPB09, and the start-up funding from University of Chinese Academy of Sciences (UCAS).



\end{document}